\documentclass[conference,letterpaper]{IEEEtran}
\IEEEoverridecommandlockouts
\usepackage{cite}
\usepackage{amsmath,amssymb,amsfonts}
\usepackage{algorithmic}
\usepackage{graphicx}
\usepackage{textcomp}
\usepackage{physics}
\usepackage{amsthm}
\usepackage{xcolor}
\usepackage{enumitem}
\usepackage{subcaption}
\usepackage{xcolor,colortbl}
\usepackage{tikz}
\usetikzlibrary{arrows.meta}
\usepackage{balance}
\usepackage{changepage}

\tikzstyle{gate}=[draw, minimum size=1em,scale=1.2]
\tikzset{meter/.append style={draw, inner sep=5, rectangle, font=\vphantom{A}, minimum width=25, line width=.4,
 path picture={\draw[black] ([shift={(.05,.15)}]path picture bounding box.south west) to[bend left=50] ([shift={(-.05,.15)}]path picture bounding box.south east);\draw[black,] ([shift={(0,.05)}]path picture bounding box.south) -- ([shift={(.15,-.05)}]path picture bounding box.north);}}}

\usepackage{qcircuit}

\newif\ifshowdiscussion
\showdiscussiontrue

\usepackage{balance}
\def\BibTeX{{\operatorname B\kern-.05em{\sc i\kern-.025em b}\kern-.08em
    T\kern-.1667em\lower.7ex\hbox{E}\kern-.125emX}}
\begin{document}

\title{On the Advantage of Irreversible Processes in Single-System Games
\\
\thanks{This work is supported by the Swiss National Science Foundation (SNF)  and the Fonds de recherche du Qu\'ebec -- Nature et technologies (FRQNT).}}

\newtheorem{definition}{Definition}
\newtheorem{proposition}{Proposition}
\newtheorem{theorem}{Theorem}
\newtheorem{lemma}{Lemma}
\newtheorem{corollary}{Corollary}
\newtheorem{observation}{Observation}
\newcommand{\VPA}{(\mathcal{V},\mathcal{P},\mathcal{A})}

\author{\IEEEauthorblockN{Xavier Coiteux-Roy and Stefan Wolf}
\IEEEauthorblockA{\textit{Faculty of Informatics},
\textit{Universit\`a della Svizzera italiana},
\textit{Lugano, Switzerland} \\
\{xavier.coiteux.roy, stefan.wolf\}@usi.ch}
}

\maketitle
\begin{abstract}
The CHSH no-signalling game studies Bell nonlocality by showcasing a gap between the win rates of classical strategies, quantum-entangled strategies, and no-signalling strategies. Similarly, the CHSH* single-system game explores the advantage of irreversible processes by showcasing a gap between the win rates of classical reversible strategies, quantum reversible strategies, and irreversible strategies. The irreversible process of erasure rules supreme for the CHSH* single-system game, but this {\it erasure advantage} does not necessarily extend to every single-system game: We introduce the {32-Game}, in which reversibility is irrelevant and only the distinction between classical and quantum operations matters. We showcase our new insight by modifying the CHSH* game to make it erasure-immune, while conserving its quantum advantage. We conclude by the reverse procedure: We tune the {32-Game} to make it erasure-vulnerable, and erase its quantum advantage in the process. The take-home message is that, when the size of the single-system is too small for Alice to encode her whole input, quantum advantage and erasure advantage can happen independently.
\end{abstract}


\section{Introduction To Single-System Games: CHSH*}
No-signalling games (sometimes called {\it nonlocal} games) demonstrate that exploiting quantum mechanics---or more precisely, quantum entanglement---provide an advantage in certain distributed-computing tasks. The CHSH game~\cite{einstein1935can,bell1964einstein} is the most well-known example of such games; the RGB no-signalling game is another very simple example~\cite{coiteux2019rgb}. For both of these games, sharing quantum entanglement allows to win with better probability than using purely classical strategies, but not with probability 1 (this was first proven by Tsirelson~\cite{cirel1980quantum}); and for both of these games, hypothetical no-signalling devices called Popescu-Rohrlich boxes~\cite{PR94} (PR-boxes, or non-local boxes) make winning with certainty possible. 

\begin{figure}[h!]
\centering \scalebox{1}{
\hspace*{0em}\begin{tikzpicture}
\node[] at (0,0) (S) {$\ket{0}$};

\node[gate] at (1.5,0) (A) {$\operatorname{A}_a$};
\node[left] at (1.5,1) (a) {$\{0,1\}\ni a \hspace{-6pt}$};

\node[gate] at (3,0) (B) {$\operatorname{B}_b$};
\node[right] at (3,1) (b) {$\hspace{-6pt}b\in \{0,1\}$};

\node[meter,scale=0.8] at (4.5,0) (M) {};
\node[anchor=west,scale=1.2] at (5.5,0) (E) {$m~\smash{ \overset{?}{=}} ~a\cdot b$};

\draw (S) -- (A) -- (B) -- (M);
\draw[double,double distance=.15em,-Implies] (M) -- (E);
\draw[double,double distance=.15em] (B) -- (b.south west);
\draw[double,double distance=.15em] (A) -- (a.south east);
\end{tikzpicture}}
\caption{CHSH* is a single-system game that exhibits a behaviour analog to the CHSH no-signalling game. Quantum reversible gates can win better than classical reversible ones, but cannot reach perfection; while the erasure gate can.}\label{CHSH*}\end{figure}
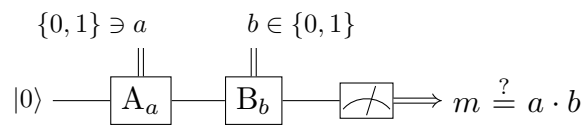

Single-system games do not study locality, but space-constrained computations. They were recently (re-)introduced\footnote{The CHSH* single-system game is a two-input read-once branching program. Branching programs are a common computational model in complexity theory~\cite{wegener2000branching}; their quantum version has been briefly studied in \cite{ablayev2005computational}.} by {\it Henaut, Catani et al.}~\cite{henaut2018tsirelson}, who reframed the standard CHSH game into a game, CHSH* (Fig.~\ref{CHSH*}), where the two players---instead of being spatially separated---are limited to deterministically applying conditional gates on a common 2-dimensional system. {\it Henaut, Catani et al.} then analyzed the best performance that can be achieved when the players are restricted to certain types of gates and found out that quantum reversible gates can do better than classical reversible gates, but cannot win with certainty---a scenario very similar to the no-signalling case, but where the advantage comes simply from the geometry of the quantum state, and not from entanglement. From their results (see some of them in the leftmost column of Tab.~\ref{table}), the question that stands out is whether the erasure gate is to the CHSH* single-system game what PR boxes are to the standard CHSH no-signalling game.

\definecolor{Gray}{gray}{0.85}
\begin{table}[htp]
\vspace{-0.5cm}
\begin{center}
\begin{tabular}{|c|c|c|c|c|}
\cline{2-5}
\multicolumn{1}{c}{} & \multicolumn{4}{|c|}{Optimal Win Rate} \\
 \hline
Gate Set & CHSH$^*$ & EI-CHSH$^*$& {32-Game}$$& {B{32-Game}}$$\\
\hline
\rowcolor{Gray}
Classical reversible & 3/4 &3/4& 7/9 & 4/5\\
\rowcolor{Gray}
Classical (ir)reversible & 1 & 3/4&7/9 & 13/15\\
Quantum reversible & 0.85 & 0.85 & 5/6& 4/5\\
Quantum (ir)reversible &  1 & --- &5/6& 13/15\\
\hline
\end{tabular}
 \begin{adjustwidth}{0.2cm}{}
\caption{We examine if having access to irreversible processes can improve Alice and Bob's---whether classical or quantum---win rates in three new single-system games (rightmost columns). For CHSH and the {B{32-Game}}, it does; while for EI-CHSH and the {32-Game}, it does not. ``$0.85$'' stands for Tsirelson's bound ($1/2+\sqrt{2}/4$).}\label{table}
\end{adjustwidth}
\end{center}
\end{table}
\vspace{-0.5cm}
To investigate this parallel, we devise a new single-system game, the {32-Game}, for which irreversible processes are not superior to reversible ones (Sec.-\ref{section2}).

We explain this difference with the CHSH* game by the fact that in the {32-Game} there exists another limitation besides reversibility---the size of the system (2-dimensional) is smaller than the size of Alice's input (a trit)---and it acts as a bottleneck that makes reversibility irrelevant. We then present {EI-CHSH*}, an erasure-immune variant of the CHSH* game, in which we artificially create this bottleneck so that irreversible processes lose their edge (Sec.~\ref{sectionEICHSH*}).

In both {32-Game} and {EI-CHSH*}, the disparity between the input size and the system size does not, however, prevent the existence of a quantum advantage.\footnote{A striking example of a qubit memory working with high-dimensional inputs is given in Ref.~\cite{ablayev2014very}, where a streaming algorithm only needs a quantum memory of $1$ qubit to compute without error a function given some promise, while, were the memory classical, $\log n$ classical bits would be necessary.}

Finally, we show that the {32-Game}'s gate hierarchy is fragile: We bias the input distributions of our first game---we rename it {B{32-Game}}---and give back an advantage to irreversible gates, while neutralizing the quantum advantage (Sec.~\ref{sectionB32}). Our results are summarized in Tab.~\ref{table}.

\section{Preliminaries: Sets of Gates}
We study the optimal strategies for two players (Alice and Bob) that are restricted to operating on a 2-dimensional system using various types of logical gates.
We ignore statistical mixture of reversible gates because---when the input distribution is fixed---drawing gates randomly following some distribution is never better than applying the best strategy from this distribution.

We study four sets of gates---classical reversible, classical (ir)reversible, quantum reversible, and quantum (ir)reversible---and use a fixed projective measurement. We represent all gates as quantum channels acting on 2-dimensional systems, using the formalism of quantum information theory (The books~\cite{nielsen2002quantum}~and~\cite{wilde2013quantum} are two excellent references). 

\paragraph{Classical Reversible}
We say the channel is classical reversible if it acts as a permutation between classical states.
A classical state is a quantum state that is diagonal in the rectilinear basis.
There are only two classical reversible gates for 2-dimensional systems: the identity $\operatorname{I}:=\begin{pmatrix}1&0\\0&1
\end{pmatrix}$, and the bit-flip $\operatorname{X}:=\begin{pmatrix}0&1\\1&0\end{pmatrix}$.

\paragraph{Quantum Reversible}
A quantum reversible channel is simply a unitary operator $\operatorname{U}$, with $\operatorname{U}^\dagger \operatorname{U}=\operatorname{U} \operatorname{U}^\dagger =\operatorname{I}$.

\paragraph{Classical (Ir)reversible}
We write (ir)reversible to designate channels that are not necessarily reversible. We build the classical (ir)reversible set of gates by adding the erasure gate to the classical reversible--gates set: The erasure gate simply outputs $0$ (or 1) no matter the input; it is an irreversible process. While such erasure is a classical operation, it can be seen as an amplitude-damping quantum channel of probability 1, and be represented by Kraus operators $K_{E1},K_{E2}:=\begin{pmatrix}1&0\\0&0\end{pmatrix},\begin{pmatrix}0&1\\0&0\end{pmatrix}$. Erasing to 1 can be obtained by erasing first to 0, and then flipping the result.

\begin{observation}\label{obs1}
When Bob applies a classical (ir)reversible gate $\operatorname{B}_b$ on the system sent by Alice (on $x_a$), it can be more intuitive to visualize him as receiving communication $x_a$ from Alice, and then using it to condition a classical reversible gate $\operatorname{B}_{x_a,b}$.
\end{observation}

\paragraph{Quantum (Ir)reversible}
Quantum (ir)reversible gates are arbitrary quantum channels and constitute the most general set of gates that we will consider. They can be represented by Kraus operators $\{K_i\}$, such that $\sum_i K_i^\dagger K_i=\operatorname{I}$, and whose action on a quantum state $\rho$ is
$\mathcal{C}(\rho):=\sum_i K_i \rho K_i^\dagger \,.$

Note that closed quantum systems follow Schr\"odinger's equation and always evolve unitarily: Genuinely irreversible quantum gates can thus physically only happen in open systems, where an external leakage of information is possible.

\paragraph{Rectilinear Measurement}
A measurement in the rectilinear basis of a state $\rho$ gives the result~1 with probability $\tr (\rho \ketbra{1})$.

\section{Quantum Advantage But No Irreversibility Advantage: The {32-Game}}\label{section2}

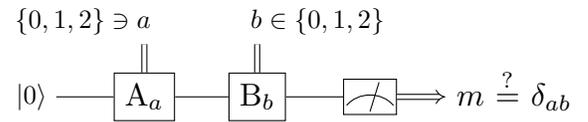
\begin{figure}[h!]
\centering

\scalebox{1}{

\tikzstyle{gate}=[draw, minimum size=1em,scale=1.2]
\tikzset{meter/.append style={draw, inner sep=5, rectangle, font=\vphantom{A}, minimum width=25, line width=.4,
 path picture={\draw[black] ([shift={(.05,.15)}]path picture bounding box.south west) to[bend left=50] ([shift={(-.05,.15)}]path picture bounding box.south east);\draw[black,] ([shift={(0,.05)}]path picture bounding box.south) -- ([shift={(.15,-.05)}]path picture bounding box.north);}}}

\hspace*{0em}\begin{tikzpicture}

\node at (0,0) (S) {$\ket{0}$};
\node[gate] at (1.5,0) (A) {$\operatorname{A}_a$};
\node[left] at (1.5,1) (a) {$\{0,1,2\} \ni a \hspace{-6pt}$};

\node[gate] at (3,0) (B) {$\operatorname{B}_b$};
\node[right] at (3,1) (b) {$\hspace{-6pt}b\in \{0,1,2\}$};

\node[meter,scale=0.8] at (4.5,0) (M) {};
\node[anchor=west,scale=1.2] at (5.5,0) (E) {$m~\smash{ \overset{?}{=}} ~\delta_{ab}$};

\draw (S) -- (A) -- (B) -- (M);
\draw[double,double distance=.15em,-Implies] (M) -- (E);
\draw[double,double distance=.15em] (B) -- (b.south west);
\draw[double,double distance=.15em] (A) -- (a.south east);

\end{tikzpicture}
}
\caption{Alice and Bob must each choose 3 two-dimensional gates. They then independently receive a random input trit and apply the corresponding gates on an initial state $\ket{0}$, which is ultimately measured in the rectilinear basis: The winning condition is that the measurement yields 1 when their inputs are identical, and that it yields 0 when their inputs are different. We vary the set of operations they can choose from.}
  \label{32game}
\end{figure}

\begin{definition}\rm
The {\it {32-Game}} is the single-system game defined in Fig.~\ref{32game} with the inputs $a,b$ drawn uniformly at random ($\forall i,j\in\{0,1,2\}^2, \operatorname{Pr}(a=i,b=j)=1/9$).
\end{definition}

\subsection{Classical Gates}
\begin{proposition}
The best classical reversible strategy for the $32$-Game wins it with probability $7/9$.
\begin{proof}
The following strategy wins with probability $7/9$ using only classical reversible gates:
$
\operatorname{A}_a:= \operatorname{X}^{a}\,,
\,\operatorname{B}_b:= \operatorname{X}^{b+1}.$
It loses only when the inputs are $(a=0,b=2)$ or $(a=2,b=0)$. Its optimality is a direct corollary of Proposition~\ref{32cproof}.
\end{proof}
\end{proposition}

\begin{proposition}\label{32cproof}
Classical (ir)reversible strategies cannot win the $32$-Game better than classical strictly reversible strategies.
\begin{proof}
We use Observation~\ref{obs1} and prove that classical (ir)reversible strategies cannot win more often than with probability $7/9$.

The state of the system after Alice's action is $x_a:=\operatorname{X}^a \ket{0}$. We view $x_a$ as 1~bit of communication from Alice to Bob. Up to relabelling of the inputs, there are two different behaviours for Alice: Either $x_0=x_1=x_2$, which is equivalent to no communication and cannot win with probability better than $2/3$, or $x_0=x_1\neq x_2$. In this case, let {$y(x_a,b):=\operatorname{B}_{x_a,b}\operatorname{X}^a \ket{0}$} be Bob's output: Then $y(x_0,0)=y(x_1,0)$ and $y(x_0,1)=y(x_1,1)$, but their respective winning conditions, $\delta_{00}\neq\delta_{10}$ and $\delta_{01}\neq\delta_{11}$, are orthogonal---Bob is sure to get at least one wrong answer for each couple---and the win rate is, therefore, at most $7/9$.\end{proof}
\end{proposition}

\subsection{Quantum Gates}
\begin{proposition}\label{qstrate}
The best quantum reversible strategy for the $32$-Game wins it with probability~$5/6$.
\begin{proof}
If we define $\operatorname{R}_{2\pi/3}:=\begin{pmatrix}\cos{\pi/3}&-i \sin{\pi/3}\\-i \sin{\pi/3}&\cos{\pi/3}\end{pmatrix}$,
the following quantum reversible strategy wins with probability $5/6$:
\begin{equation*}
\operatorname{A}_a:= \operatorname{X} \cdot \operatorname{R}_{2\pi/3}^{a}\,,
\,\operatorname{B}_b:= \operatorname{R}_{2\pi/3}^{2b}\,.
\end{equation*}
Note that $\operatorname{A}_a\cdot \operatorname{B}_{a}=\operatorname{I}$; $\operatorname{A}_a\cdot \operatorname{B}_{\neg a}\in \{\operatorname{R}_{2\pi/3},\operatorname{R}_{2\pi/3}^{2} \}$ and  $\tr (\operatorname{R}_{2\pi/3} \ketbra{0})=\tr (\operatorname{R}_{2\pi/3}^2 \ketbra{0})=1/4$.

Its optimality is a direct corollary of Proposition~\ref{32qproof}.
\end{proof}
\end{proposition}

\begin{definition}\rm\label{discr-exp}
The {\it discrimination experiment} corresponding to the {$32$-} single-system game is the following: 0) Alice and Bob agree on three quantum states $\rho_0,\rho_1,\rho_2$. 1) Alice is given her random input $a\in\{0,1,2\}$, and sends to Bob $\rho_a$. 2) Bob is given his random input $b\in\{0,1,2\}$, and needs to guess, using any means allowed by quantum mechanics, whether Alice sent him $\rho_{b}$ or not. Alice and Bob play together in trying to minimize Bob's probability of error.
\end{definition}

\begin{lemma}\label{discr-quant}
The probability of winning the discrimination experiment bounds from above the probability of winning the 32- single-system game using a quantum (ir)reversible strategy.
\begin{proof}
Any quantum (ir)reversible strategy for the single-system game can be turned into a strategy for the discrimination experiment that wins with the same probability; it thus cannot be better.

The strategy for the discrimination experiment is the following: Alice sends the conditional state $\operatorname{A}_a \ket{0}$ to Bob; Bob applies the quantum channel~$\operatorname{B}_b$ and then measures in the rectilinear basis. On outcome 0 he guesses that Alice sent $\rho_{b}$, and on outcome 1 he guesses otherwise.
\end{proof}
\end{lemma}

\begin{proposition}\label{32qproof}
Quantum (ir)reversible strategies cannot win the $32$-Game better than quantum strictly reversible strategies.

\begin{proof}
We bound from above the probability of winning the discrimination-experiment scenario, and the conclusion then follows from Lemma~\ref{discr-quant}.

The minimal-error measurement for distinguishing two arbitrary quantum states $\rho$ and $\sigma$, of respective prior probabilities $p$ and $q$, was characterized by Helstrom~\cite{helstrom1976quantum}. It gives a tight bound on the distinguishability success rate:
\begin{equation*}
p_{\rm guess}\le \frac{1}{2} + \frac{1}{2} \norm{p \rho - q \sigma}_1\,, {\rm~where~} \norm{A}_1:=\tr \sqrt{AA^\dagger}\,.
\end{equation*}

We apply it to the {$32$-Game} discrimination-experiment scenario:
\begin{align}
p_{\rm guess}&\le\frac{1}{2} + \frac{1}{6} \sum_{i=0}^2 \norm{ \frac{1}{3}  \rho_i -  \frac{2}{3} \frac{\rho_{i+1}+\rho_{i+2}}{2}}_1\nonumber \\
&=\frac{1}{2} + \frac{1}{18} \sum_{i=0}^2 \norm{ \rho_i -  \rho_{i+1}-\rho_{i+2}}_1\,, \label{tobebounded}
\end{align}
where the sums in the indices are understood to be modulo~3.

To bound this quantity from above, we start by observing that any density matrix can be represented by a real vector $\vec{v}_i:=x_i \hat{x}+y_i\hat{y}+z_i\hat{z}$ on or inside the Bloch sphere (meaning $\norm{\vec{v}_i}_2\le 1$):
\begin{equation*}
\rho_i=(\operatorname{I}+x_i \sigma_x +y_i\sigma_y+z_i \sigma_z  )/2\,.
\end{equation*}
We introduce the changes of variables
\begin{equation*}
\vec{r}_i:=\vec{v}_i-\vec{v}_{i+1}-\vec{v}_{i+2}\,.
\end{equation*}
Eq.~\ref{tobebounded} becomes 
\begin{equation}
p_{\rm guess}\le\frac{1}{2} + \frac{1}{18} \sum_{i=0}^2 \norm{ \overbrace{\frac{-\operatorname{I} + r_{i,x} \sigma_x +r_{i,y} \sigma_y+r_{i,z} \sigma_z}{2}}^{:=A_i}}_1\,. \label{inject2}
\end{equation}

To evaluate $\norm{A_i}_1$, we note that $-A_i$ is a hermitian matrix with eigenvalues $(1\pm \norm{\vec{r}_i}_2)/2$,
and that, therefore,
\begin{align}
\norm{A_i}_1=\tr\sqrt{A_iA_i^\dagger}&=(\left|1+\norm{\vec{r}_i}_2\right|+\left|1-\norm{\vec{r}_i}_2\right|)/2\label{eq10}\\
&=   \begin{cases}
   \norm{\vec{r}_i}_2 &\text{if $\norm{\vec{r}_i}_2\ge1$} \nonumber \\
   1 &\text{if $\norm{\vec{r}_i}_2<1$}\,.\nonumber
   \end{cases}
\end{align}

We now separate the analysis into 4 different cases.

\begin{definition}\rm
We separate the strategy distributions for the {32-Game} discrimination experiment using the parameter
\begin{equation*}
D_{\rm max}:=\# \{ i {\rm~s.t.~}\norm{A_i}_1=1\}\,. \label{DSdef}
\end{equation*}
\end{definition}

$D_{\rm max}$ corresponds to the maximum number of inputs $b=i$ for which Bob could completely ignore Alice's action and guess according to the highest prior; this is what happens effectively when for a certain input Bob uses an erasure gate: He wins with conditional probability $\max(p_i,q_i)=2/3$ no matter Alice's behaviour.

\paragraph{Case $D_{\rm max}=0$}
We use the Cauchy-Schwarz inequality, and then the three length constraints, to obtain (note that $\norm{\vec{v}_i}_2=:\vec{v}_i^2$ in the dot-product notation)
\begin{align}
&\sum_{i=0}^{2}\norm{A_i}_1=\norm{\vec{r}_0}_2+\norm{\vec{r}_1}_2+\norm{\vec{r}_2}_2 \label{eq13}\\&\le
\sqrt{3} \norm{\frac{\strut}{\strut} \norm{\vec{r}_0}_2+\norm{\vec{r}_1}_2+\norm{\vec{r}_2}_2 }_2\nonumber\\
&=\sqrt{3}\sqrt{3(\vec{v}_0^2+\vec{v}_1^2+\vec{v}_2^2) -2(\vec{v}_0\cdot\vec{v}_1+\vec{v}_1\cdot\vec{v}_2+\vec{v}_0\cdot\vec{v}_2)}\nonumber\\
&\le\sqrt{3}\sqrt{ 9 -2(\vec{v}_0\cdot\vec{v}_1+\vec{v}_1\cdot\vec{v}_2+\vec{v}_0\cdot\vec{v}_2)}\,.\nonumber
\end{align}

Our task is now to bound from below
\begin{align}
\vec{v}_0\cdot\vec{v}_1+\vec{v}_1\cdot\vec{v}_2+\vec{v}_0\cdot\vec{v}_2 \tag{minimize}\\
\forall i, \vec{v}_i^2\le1 \tag{constraints}\,.
\end{align}

We use a Lagrangian multipliers method (the $s_i$ are slack constraints):
\begin{equation*}
\mathcal{L}(\{\vec{v}_i,\lambda_i,s_i\}_{i=0}^{i=2})=
\vec{v}_0\cdot\vec{v}_1+\vec{v}_1\cdot\vec{v}_2+\vec{v}_0\cdot\vec{v}_2  + \sum_{i=0}^2 \lambda_i (\vec{v}_i^2+s_i^2-1)\,.
\end{equation*}

The set of vectors minimizing our function include at least one vector that saturates the unit-length constraint\footnote{That is because if none of the length constraints were saturated, a homothety could amplify the solution (its value is negative) and minimize further the function.}; without a loss of generality, and invoking the spherical symmetry, let us say that it is $\vec{v}_0=\hat{x}$.

We pose $\grad\mathcal{L}=0$ and obtain 15 scalar equalities. Examples are
\begin{align*}
\pdv{\mathcal{L}}{y_0}&=0 \iff y_1+y_2=0\,,\\
\pdv{\mathcal{L}}{y_1}&=0 \iff y_2 +2\lambda_1 y_1=0\,,\\
\pdv{\mathcal{L}}{x_0}&=0 \iff x_1 + x_2 + 2 \lambda_0 = 0\,,\\
\pdv{\mathcal{L}}{x_1}&=0 \iff 1+x_2 +2\lambda_1 x_1=0\,,
\end{align*}
Comparing the first two equations, and then the next two, we find that either $\lambda_0=\lambda_1=1/2$, or $y_1=y_2=0$. Similarly, if we were to take in the $\hat{z}$ direction the analogues of the first two equations, we would conclude that either $\lambda_0=\lambda_1=1/2$, or $z_1=z_2=0$. This implies that if $\lambda_0\neq1/2$, all vectors are co-linear (in the $\hat{x}$ direction) and the strategy is classical, but then they are of no interest since classical strategies cannot win better than 7/9 (Prop.~\ref{32cproof}).

We thus assume $\lambda_0=1/2$, and go back to $\grad\mathcal{L}=0$. We develop the vectorial equality
\begin{equation*}
\pdv{\mathcal{L}}{\vec{v}_0}=0
\iff \vec{v}_0 + \vec{v}_1 + \vec{v}_2 = 0\,,
\end{equation*}
square it,
\begin{equation*}
\left(\vec{v}_0 + \vec{v}_1 + \vec{v}_2 \right)^2= 0\,,
\end{equation*}
and expand it, in conjunction with the length constraints, to conclude finally that
\begin{equation}
\vec{v}_0\cdot\vec{v}_1+\vec{v}_1\cdot\vec{v}_2+\vec{v}_0\cdot\vec{v}_2 = - (\vec{v}_0^2+\vec{v}_1^2+\vec{v}_2^2)/2 \ge -3/2 \,. \tag{solution}
\end{equation}

Injecting this bound back into Eq.~\ref{eq13}, and then Eq.~\ref{inject2}, we find that $p_{\rm guess}\le 5/6$ for the case $D_{\rm max}=0$. 

The other cases do not violate this bound as the following crude inequalities show.
\paragraph{Case $D_{\rm max}=1$}
Without losing generality, we pose $\norm{A_0}_1=1$. We re-write Eq.~\ref{eq13} (the previous case achieves 6).
\begin{align*}
&\sum_{i=0}^{2}\norm{A_i}_1=1+\norm{\vec{r}_1}_2+\norm{\vec{r}_2}_2 \\&\le
1+\sqrt{2} \cdot \norm{\frac{\strut}{\strut} \norm{\vec{r}_0}_2+\norm{\vec{r}_1}_2+\norm{\vec{r}_2}_2 }_2\\
&\le 1+\sqrt{2}\sqrt{12}\approx 5.9\,.
\end{align*}
\paragraph{Case $D_{\rm max}=2$}
Without losing generality, we pose $\norm{A_0}_1=\norm{A_1}_1=1$. Then Eq.~\ref{eq13} becomes
\begin{align*}
&\sum_{i=0}^{2}\norm{A_i}_1=2+\norm{\vec{r}_2}_2 \le5\,.
\end{align*}
\paragraph{Case $D_{\rm max}=3$}
Finally, $\sum_{i=0}^{2}\norm{A_i}_1=3$.

This proves that none of the 4~cases could allow a better probability of winning than $p_{\rm guess}\le 5/6$, the one achieved in Prop.~\ref{qstrate}, and which is then optimal.
\end{proof}
\end{proposition}

\section{Generalizing the insight: EI-CHSH*}\label{sectionEICHSH*}
\subsection{When Are Classical (Ir)reversible Gates Optimal?}
\begin{proposition}\label{suff}
For any 2-player single-system game, classical (ir)reversible strategies are as good as quantum (ir)reversible strategies if the dimension of the system\footnote{We had defined so far the system to be 2-dimensional.} is at least as large as the size of Alice's input.
\begin{proof}
It follows from Observation~\ref{obs1}: If Alice can communicate to Bob her full input, Bob can make sure to produce a winning output whenever one would be possible.
\end{proof}
\end{proposition}
This condition, which is respected in the CHSH* game, delimitates when the parallel between the set of (ir)reversible gates and the class of no-signalling correlations is warranted. In the framework of sequential transformation contextuality (as introduced in Ref.~{\cite{mansfield2018quantum}}), restrictions on the system size are not reflected at the ontological level, and the condition in Prop.~\ref{suff} is, therefore, always fulfilled.

\subsection{When is Irreversibility Irrelevant for Classical Gates?}

\begin{proposition}\label{general_statement}
For any binary-output 2-player single-system game, a necessary condition for classical (ir)reversible processes to provide an advantage over classical reversible processes is that
\begin{equation}
\exists b \mbox{\rm~s.t.~} \sum_a p_{a,b}  W_{a,b}^{(0)} \neq \sum_a p_{a,b}  W_{a,b}^{(1)} \,,\label{necc}
\end{equation}
where $p_{a,b}$ is the prior of the inputs, and
${W_{a,b}^{(o)}=\begin{cases}
   1 &\text{if output $o$ wins on inputs $(a,b)$}\,, \\
   0 &\text{otherwise}\,.
   \end{cases}}$
\begin{proof}
The negation of Eq.~\ref{necc} implies that for all inputs $b$ of Bob, erasing to zero or erasing to one wins with the same probability (when erasing a two-dimensional system, Bob effectively decides to ignore Alice's input). In such cases, Bob could as well choose at random $\operatorname{I}$ or $\operatorname{X}$. Therefore, at least one of $\operatorname{I}$ and $\operatorname{X}$ is at least as good as erasing. The latter cannot, therefore, provide any advantage over reversible operations.\end{proof}
\end{proposition}

As shown by the {32-Game} with uniformly distributed inputs---where this condition is satisfied but there is still no erasure advantage---this condition is necessary, but not sufficient. 

It is open how Prop.~\ref{general_statement} extends to the quantum case, or how to generalize it to higher-dimension outputs.

\subsection{The Erasure-Immune CHSH* Game}
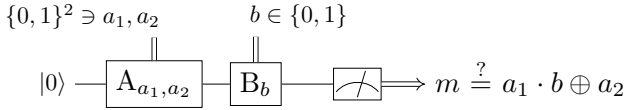
\begin{figure}[h!]
\vspace{-0.5cm}
\centering \scalebox{0.9}{
\hspace*{0em}\begin{tikzpicture}
\node[] at (0,0) (S) {$\ket{0}$};

\node[gate] at (1.5,0) (A) {$\operatorname{A}_{a_1,a_2}$};
\node[left] at (1.5,1) (a) {$\{0,1\}^2\ni a_1,a_2 \hspace{-6pt}$};

\node[gate] at (3,0) (B) {$\operatorname{B}_b$};
\node[right] at (3,1) (b) {$\hspace{-6pt}b\in \{0,1\}$};

\node[meter,scale=0.8] at (4.5,0) (M) {};
\node[anchor=west,scale=1.2] at (5.5,0) (E) {$m~\smash{ \overset{?}{=}} ~a_1\cdot b \oplus a_2$};

\draw (S) -- (A) -- (B) -- (M);
\draw[double,double distance=.15em,-Implies] (M) -- (E);
\draw[double,double distance=.15em] (B) -- (b.south west);
\draw[double,double distance=.15em] (A) -- (a.south east);
\end{tikzpicture}}
\caption{In this erasure-immune variant of the CHSH* game, classical irreversible processes do not win better than reversible ones, but the quantum advantage remains.}

  \label{immune_CHSH*}
\end{figure}

We illustrate Prop.~\ref{general_statement} by modifying the CHSH* game as to remove its erasure advantage, while keeping the quantum-unitary advantage. In the variant, Alice is given a second output that inverts the winning condition with probability $1/2$.\footnote{Note that, differently for example from the secure delegated computing case of Ref.~\cite{dunjko2014quantum}, Alice is here working with Bob and is not purposely hiding her input from him. Her input is simply too large for the size 2 of the single system.}
\begin{definition}\rm
The {\it erasure-immune CHSH*} single-system game (EI-CHSH*) is defined in Fig.~\ref{immune_CHSH*}. The inputs are selected uniformly at random.
\end{definition}

\begin{proposition}
Classical (ir)reversible gates are not better than classical strictly reversible ones in the erasure-immune CHSH* game: They win with at most probability $3/4$; while a quantum-unitary strategy can reach Tsirelson's bound $(\approx 0.85)$. These bounds are tight.
\begin{proof}
Any reversible strategy for the CHSH* game---whether classical or quantum---can be turned into a strategy for the erasure-immune CHSH* game with the same winning rate, and {\it vice versa}: Alice simply needs to apply a $\operatorname{X}^{a_2}$ gate at the very beginning of the circuit, effectively turning, when $a_2=1$, the $\ket{0}$ initial state into $\ket{1}$ (this works because for 2-dimensional reversible gates, the transition rates $\ket{0} \leftrightarrow \ket{1}$ of any strategy are symmetric).

The absence of advantage of classical (ir)reversible processes is a direct consequence of Prop.~\ref{general_statement}.
\end{proof}
\end{proposition}

\section{Recovering the Irreversibility Advantage: The {B{32-Game}}}\label{sectionB32}
We modify the {32-Game} as to make irreversibility relevant again, and lose the quantum advantage in the process.
\begin{definition}\rm
We call {\it Biased {32-Game}} ({B{32-Game}}) the same single-system game that is defined in Fig.~\ref{32game}, but with the biased input distribution
\begin{equation*}
\forall i,j: p_{a=i,b=j}=
\begin{cases}
   1/15 &\text{if $i=j$}\,, \\
   2/15 &\text{otherwise}\,.
   \end{cases}
\end{equation*}
 \end{definition}

\begin{proposition}
The best classical reversible strategy for the biased $32$-Game wins it with probability $4/5$.
\begin{proof}
For each of their inputs, Alice and Bob can choose between two operations ($\operatorname{I}$ or $\operatorname{X}$). This make for a total of only $2^6$ classical reversible strategies; an exhaustive search reveals none of them win more than with probability $4/5$.
\end{proof}
\end{proposition}
\begin{proposition}
The best quantum reversible strategy for the biased $32$-Game also wins it with probability $4/5$.
\begin{proof}
Proof omitted. It can be proven using the semi-definite programming technique developed in Ref.~\cite{wehner2006tsirelson}, and as it is used in Ref.~\cite{coiteux2019rgb}, to optimize the winning probability of a no-signalling game version of $32$-Game, and then apply the reduction from the no-signalling game to the single-system game as it is done in Lemma~1 of Ref.~\cite{henaut2018tsirelson}.\end{proof}
\end{proposition}

\begin{proposition}\label{32qproof2}
The best classical (ir)reversible strategy for the biased $32$-Game wins it with probability $13/15$.

\begin{proof}
The following classical (ir)reversible strategy is sufficient to win with probability $p_{\rm win} = 1/3 + 2/3 \cdot 4/5 = 13/15$:
Alice applies $\operatorname{X}$ if $a = 0$, and $\operatorname{I}$ otherwise; while Bob applies $\operatorname{I}$ if $b=0$, and erases Alice's bit otherwise. 

To show it is optimal, we prove that not even quantum (ir)reversible strategies can do better. We use the same discrimination-experiment technique as in Section~\ref{section2}, but skip over the details.
Eq.~\ref{inject2} and  Eq.~\ref{eq10} become 
\begin{align*}
p_{\rm guess}&\le\frac{1}{2} + \frac{1}{30} \sum_{i=0}^2 \norm{ \overbrace{\frac{-3 + r_{i,x} \sigma_x +r_{i,y} \sigma_y+r_{i,z} \sigma_z}{2}}^{:=A_i}}_1\,{\rm and} \\
\norm{A_i}_1&=\tr\sqrt{A_iA_i^\dagger}= \begin{cases}
   \norm{\vec{r}_i}_2 &\text{if $\norm{\vec{r}_i}_2\ge3$} \\
   1 &\text{if $\norm{\vec{r}_i}_2<3$}\,.
   \end{cases}
\end{align*}

We again analyze the four cases $D_{\rm max}=0,1,2,3$.
\paragraph{Case $D_{\rm max}=0$}
Eq.~\ref{eq13} becomes 
\begin{align*}
&\sum_{i=0}^{2}\norm{A_i}_1=\norm{\vec{r}_0}_2+\norm{\vec{r}_1}_2+\norm{\vec{r}_2}_2\\
&\le\sqrt{3}\sqrt{ 3\cdot (3+3+3)} = 9;
\end{align*}
all crossed terms disappeared. This bound implies a maximal winning rate of only
$p_{\rm guess}\le\frac{1}{2} + \frac{9}{30} =4/5\,,$
which is the same as Bob simply guessing $(\rho_{b+1}+\rho_{b+2})/2$ without looking at what Alice sent him.
\paragraph{Case $D_{\rm max}=3$}
$\sum_{i=0}^{2}\norm{A_i}_1=9$.

\paragraph{Case $D_{\rm max}=2$}
$\sum_{i=0}^{2}\norm{A_i}_1=6+\norm{\vec{r}_2}_2 \le 11\,;$
this bound corresponds to the winning-rate $p_{\rm guess}=13/15$ of the classical reversible strategy we mentioned previously (it uses 2 conditional-erasure gates).

\paragraph{Case $D_{\rm max}=1$}
Finally,

$\sum_{i=0}^{2}\norm{A_i}_1=3+\norm{\vec{r}_1}_2+\norm{\vec{r}_2}_2 \le
3+\sqrt{2} \sqrt{27} <11\,.\qedhere$

\end{proof}
\end{proposition}

The (ir)reversible classical strategy is optimal:

\begin{corollary}
Quantum (ir)reversible processes cannot win the biased {32-Game} better than classical (ir)reversible ones.\end{corollary}

\section*{Acknowledgment}
The authors thank the anonymous referees, and Anna Pappa, for helpful comments and discussions. Furthermore, Xavier thanks Lorenzo Catani for having presented him his poster~\cite{postercatani} in Zurich, Pierre McKenzie for having shown him permutation branching programs, and Arne Hansen for his many TikZ advices.

\newpage
\bibliographystyle{IEEEtranbetter}
\bibliography{IEEEabrv,tetra}

\begin{thebibliography}{10}
\providecommand{\url}[1]{#1}
\csname url@samestyle\endcsname
\providecommand{\newblock}{\relax}
\providecommand{\bibinfo}[2]{#2}
\providecommand{\BIBentrySTDinterwordspacing}{\spaceskip=0pt\relax}
\providecommand{\BIBentryALTinterwordstretchfactor}{4}
\providecommand{\BIBentryALTinterwordspacing}{\spaceskip=\fontdimen2\font plus
\BIBentryALTinterwordstretchfactor\fontdimen3\font minus
  \fontdimen4\font\relax}
\providecommand{\BIBforeignlanguage}[2]{{%
\expandafter\ifx\csname l@#1\endcsname\relax
\typeout{** WARNING: IEEEtran.bst: No hyphenation pattern has been}%
\typeout{** loaded for the language `#1'. Using the pattern for}%
\typeout{** the default language instead.}%
\else
\language=\csname l@#1\endcsname
\fi
#2}}
\providecommand{\BIBdecl}{\relax}
\BIBdecl

\bibitem{einstein1935can}
A.~Einstein, B.~Podolsky, and N.~Rosen, ``Can quantum-mechanical description of
  physical reality be considered complete?'' \emph{Physical Review}, vol.~47,
  no.~10, p.~777, 1935.

\bibitem{bell1964einstein}
J.~S. Bell, ``On the {E}instein {P}odolsky {R}osen paradox,'' \emph{Physics
  Physique Fizika}, vol.~1, no.~3, p.~195, 1964.

\bibitem{coiteux2019rgb}
X.~Coiteux-Roy and C.~Cr{\'e}peau, ``The {RGB} no-signalling game,'' in
  \emph{14th Conference on the Theory of Quantum Computation, Communication and
  Cryptography (TQC 2019)}.\hskip 1em plus 0.5em minus 0.4em\relax Schloss
  Dagstuhl-Leibniz-Zentrum fuer Informatik, 2019.

\bibitem{cirel1980quantum}
B.~S. Cirel'son, ``Quantum generalizations of bell's inequality,''
  \emph{Letters in Mathematical Physics}, vol.~4, no.~2, pp.~93--100, 1980.

\bibitem{PR94}
\BIBentryALTinterwordspacing
S.~Popescu and D.~Rohrlich, ``Quantum nonlocality as an axiom,''
  \emph{Foundations of Physics}, vol.~24, no.~3, pp.~379--385, 1994. [Online].
  Available: \url{http://dx.doi.org/10.1007/BF02058098}
\BIBentrySTDinterwordspacing

\bibitem{wegener2000branching}
I.~Wegener, \emph{Branching programs and binary decision diagrams: theory and
  applications}.\hskip 1em plus 0.5em minus 0.4em\relax SIAM, 2000, vol.~4.

\bibitem{ablayev2005computational}
F.~Ablayev, A.~Gainutdinova, M.~Karpinski, C.~Moore, and C.~Pollett, ``On the
  computational power of probabilistic and quantum branching program,''
  \emph{Information and Computation}, vol.~203, no.~2, pp.~145--162, 2005.

\bibitem{henaut2018tsirelson}
L.~Henaut, L.~Catani, D.~E. Browne, S.~Mansfield, and A.~Pappa, ``Tsirelson's
  bound and {L}andauer's principle in a single-system game,'' \emph{Physical
  Review A}, vol.~98, no.~6, p.~060302, 2018.

\bibitem{ablayev2014very}
F.~Ablayev, A.~Gainutdinova, K.~Khadiev, and A.~Yakary{\i}lmaz, ``Very narrow
  quantum obdds and width hierarchies for classical obdds,'' in
  \emph{International Workshop on Descriptional Complexity of Formal
  Systems}.\hskip 1em plus 0.5em minus 0.4em\relax Springer, 2014, pp.~53--64.

\bibitem{nielsen2002quantum}
M.~A. Nielsen and I.~Chuang, ``Quantum computation and quantum information,''
  2002.

\bibitem{wilde2013quantum}
M.~M. Wilde, \emph{Quantum information theory}.\hskip 1em plus 0.5em minus
  0.4em\relax Cambridge University Press, 2013.

\bibitem{helstrom1976quantum}
C.~W. Helstrom, \emph{Quantum detection and estimation theory}.\hskip 1em plus
  0.5em minus 0.4em\relax Academic press, 1976.

\bibitem{mansfield2018quantum}
S.~Mansfield and E.~Kashefi, ``Quantum advantage from sequential-transformation
  contextuality,'' \emph{Physical review letters}, vol.~121, no.~23, p.~230401,
  2018.

\bibitem{dunjko2014quantum}
V.~Dunjko, T.~Kapourniotis, and E.~Kashefi, ``Quantum-enhanced secure delegated
  classical computing,'' \emph{arXiv preprint arXiv:1405.4558}, 2014.

\bibitem{wehner2006tsirelson}
S.~Wehner, ``Tsirelson bounds for generalized clauser-horne-shimony-holt
  inequalities,'' \emph{Physical Review A}, vol.~73, no.~2, p.~022110, 2006.

\bibitem{postercatani}
L.~Henaut, L.~Catani, D.~E. Browne, S.~Mansfield, and A.~Pappa,
  ``Irreversibility and non-classicality in a single system game,''
  \emph{Poster presented at the Solstice of Foundation conference in Zurich},
  2019.

\end{thebibliography}

\end{document}